\documentclass[preprint,pra,showpacs,nofootinbib]{revtex4}
\usepackage{graphicx}
\usepackage{float,epsfig}
\usepackage{dcolumn}
\usepackage{bm}
\usepackage{graphicx}
\usepackage{bm}
\usepackage{amsmath,amssymb,amsthm}
\usepackage[colorlinks=true,linkcolor=blue]{hyperref}
\begin{document}
\title{\bf The Unruh effect and entanglement generation for
 accelerated atoms  near a reflecting boundary}
\author{Jialin Zhang and Hongwei Yu\footnote{Corresponding author}}
\affiliation{Department of Physics and Institute of  Physics,\\
Hunan Normal University, Changsha, Hunan 410081, China}


\begin{abstract}

We study, in the framework of open systems, the entanglement
generation of two independent uniformly accelerated atoms in
interaction with the vacuum fluctuations of massless scalar fields
subjected to a reflecting plane boundary. We demonstrate that, with
the presence of the boundary, the accelerated atoms exhibit distinct
features from static ones in a thermal bath at the corresponding
Unruh temperature in terms of the entanglement creation at the
neighborhood of the initial time. In this sense, accelerated atoms
in vacuum do not necessarily have to behave as if they were static
in a thermal bath at the Unruh temperature.

\end{abstract}
 \maketitle
\section{Introduction}

It is now well established that uniformly accelerated
observers(atoms) perceive as a thermal bath of particles at a
temperature proportional to  the proper acceleration what an
inertial observer sees as a vacuum~\cite{Fulling,unruh1,unruh2}.
 This result is known as the Unruh effect  and it seems to  imply
that the accelerating detectors (atoms) may be viewed as an open
system, i.e., a system immersed in an external thermal bath.

On the other hand,  quantum entanglement  has been recognized as a
unique quantum resource whose production can be employed for
computational and communication purposes, such as quantum
communication~\cite{Bennett}, quantum
teleportation~\cite{telportation}, quantum
cryptography~\cite{cryptography} and so on. The relationship between
entanglement and environment is an intriguing issue in the
discussions for the essence of entanglement. In this regard, it is
known that an environment usually leads to decoherence and noise,
which may cause entanglement that might have been created before to
disappear. However, in certain circumstances, the environment, such
as thermal baths belonging to a specific class, may enhance
entanglement rather than destroying
it~\cite{pr1,pr2,pr3,pr4,pr5,pr6}. The reason is that an external
environment can also provide an indirect interaction between
otherwise totally uncoupled subsystems through correlations that
exist.

Therefore, a question arises naturally as to whether entanglement
can be produced for independent accelerated atoms in vacuum, as
entanglement generation through the action of an external thermal
bath has been shown to occur. Although not directly coupled, a sea
of vacuum fluctuations of  external fields through which atoms move
may provide an indirect interaction to generate entanglement among
them. Consequently, there arises a new possibility for an
experimental test of the Unruh effect by using appropriate quantum
optics devices to detect the entanglement generated by the uniform
acceleration.

Recently, entanglement generation
 for two, independent uniformly accelerating
two-level atoms with vanishing separation interacting with a set of
scalar fields in vacuum has been examined, by Benatti et
al~\cite{Benatti1},  in the framework of open systems. In the weak
coupling limit, the completely positive dynamics for the atoms as a
subsystem has been derived by tracing over the field degrees of
freedom~\cite{Benatti1}, and there it has been shown that the
asymptotic equilibrium state of the atoms turns out to be entangled
even if the initial state is separable. Similar results have been
obtained  for atoms immersed in a thermal bath of scalar particles
at a finite temperature~\cite{Benatti2}, where the separation
between atoms is allowed to be nonzero. It is found there that for
any fixed, finite separation, there always exists a temperature
below which entanglement generation occurs as soon as time starts to
become nonzero and for the vanishing separation the entanglement
thus generated persists even in the late-time asymptotic equilibrium
state. Recently,  the problem has been further investigated assuming
the presence of a reflecting boundary for the scalar fields which
modifies the quantum fluctuations of fields~\cite{zhjl}. This
modification, which alters the field correlation functions that
characterize the fluctuations of fields, will presumably affect the
entanglement generation.~\footnote{Let us note that other novel
effects that arise from the modification include (but not limited
to) the Casimir effect~\cite{cas}, the light-cone fluctuations when
gravity is quantized~\cite{YU}, the Brownian (random) motion of test
particles in an electromagnetic vacuum~\cite{yu2}, and the
modification for the radiative properties of uniformly accelerated
atoms~\cite{Audretsch,yu3}.} In fact, it has been demonstrated that
the presence of the boundary may play a significant role in
controlling the entanglement creation in some circumstances and the
new parameter, the distance of the atoms from the boundary besides
the bath temperature and the separation of the atoms, gives one more
freedom in controlling the entanglement generation~\cite{zhjl}.

 In present paper, we are concerned with the entanglement generation of two mutually independent,
uniformly accelerated  two-level atoms interacting with a set of
massless scalar fields in the presence of a perfectly reflecting
boundary. At the first glance, one may expect the same results as
that for the case of a thermal bath (at the Unruh temperature
proportional to the acceleration) with a boundary. However, with the
help of the master equation that describes the evolution of the open
system (atoms plus external thermal fields) in time, we show, to the
contrary, that, with the presence of the boundary, the subsystem of
the uniformly accelerated atoms may behave quite differently, in
terms of entanglement generation at a neighborhood of the initial
time $t=0$, from that of static atoms immersed in a thermal bath at
the Unruh temperature.

\section{the Master Equation}
In this section, we establish the basic formalism using
 some well-known techniques for
 open quantum systems to analyze a system of two independent
uniformly accelerated two-level atoms in weak interaction with a set
of massless quantum scalar fields in the presence of a reflecting
boundary.
  We assume that the reflecting boundary for the scalar fields
is located at $z=0$ in space and  one atom is placed at point ${\bf
x}_1$ and the other at ${\bf x}_2$, with $L$ being the spatial
separation between them. Without loss of generality, the total
Hamiltonian  for the complete system (atoms+external fields) has the
form $H=H_s+H_\phi+ \lambda\;H'\;$, where $H_s$ is the Hamiltonian
of the  atom,
\begin{equation}\label{hs}
H_s=H_s^{(1)}+H_s^{(2)},\ \ H_s^{(\alpha)}=\omega\,
n_i\sigma_i^{(\alpha)}/2, \;(\alpha=1,2), \ \
\sigma_i^{(1)}=\sigma_i\otimes{\sigma_0},\ \
\sigma_i^{(2)}={\sigma_0}\otimes\sigma_i\;.
\end{equation}
Here, the $\sigma_i,\, (i=1,2,3)$  are the Pauli matrices,
$\sigma_0$ the $2\times2$ unit matrix, $\mathbf{n} =(n_1,n_2,n_3)$
a unit vector, $\omega$ the energy level spacing, and summation over
repeated index is implied. $H_\phi$ is the standard Hamiltonian of
massless, free scalar fields, details of which is not very relevant
here and $H'$ is the interaction Hamiltonian of the atoms with the
external scalar fields which is assumed to be weak
\begin{equation}
H'=\sum_{\tau=0}^{3}[(\sigma_{\tau} \otimes
\sigma_0)\Phi_{\tau}(t,{\bf {x}}_1)+(\sigma_{0} \otimes
\sigma_{\tau})\Phi_{\tau}(t,{\bf {x}}_2)]\;.
\end{equation}
 In the limit of weak-coupling, the reduced density  is found to obey
an equation in the Kossakowski-Lindblad form~\cite{Lindblad,pr5}
\begin{equation}
\frac{\partial\rho(t)}{ \partial {t}}= -i \big[H_{\rm eff},\,
\rho(t)\big]
 + {\cal L}[\rho(t)]\ ,
\label{master}
\end{equation}
where
\begin{equation}
{\cal L}[\rho]=\frac{1}{2} \sum_{\alpha,\beta=1}^2
C_{ij}^{(\alpha\beta)}\big[2\,
\sigma_j^{(\beta)}\rho\,\sigma_i^{(\alpha)}
-\sigma_i^{(\alpha)}\sigma_j^{(\beta)}\, \rho
-\rho\,\sigma_i^{(\alpha)}\sigma_j^{(\beta)}\big]\ .
\label{lindblad2}
\end{equation}
The  matrix $C_{ij}^{(\alpha\beta)}$ and $H_{\rm eff}$ are
determined by the correlation functions
\begin{equation}
G_{ij}^{(\alpha\beta)}(t-t')=\langle0|\Phi_i(t,{\bf
x}_{\alpha})\Phi_j(t',{\bf x}_{\beta})|0 \rangle\label{twogreen}\;.
\end{equation}
The  Fourier and Hilbert transforms of the correlation function
$G_{ij}^{(\alpha\beta)}$
 read respectively
\begin{equation}
{\cal G}_{ij}^{(\alpha\beta)}(\lambda)=\int_{-\infty}^{\infty} dt \,
e^{i{\lambda}t}\, G_{ij}^{(\alpha\beta)}(t)\; , \label{fourierG}
\end{equation}
\begin{equation}
{\cal K}_{ij}^{(\alpha\beta)}(\lambda)=\int_{-\infty}^{\infty} dt \,
{\rm sign}(t)\, e^{i{\lambda}t}\, G_{ij}^{(\alpha\beta)}(t)=
\frac{P}{\pi i}\int_{-\infty}^{\infty} d\omega\ \frac{ {\cal
G}_{ij}^{(\alpha\beta)}(\omega) }{\omega-\lambda} \;. \label{kij}
\end{equation}

One can show that the  Kossakowski matrix $C_{ij}^{(\alpha\beta)}$
can be written explicitly as
\begin{equation}
C_{ij}^{(\alpha\beta)}=\sum_{\xi=+,-,0} {\cal
G}_{kl}^{(\alpha\beta)}(\xi\omega)\, \psi_{ki}^{(\xi)}\,
\psi_{lj}^{(-\xi)}\;  \label{cij-2}
\end{equation}
where
\begin{equation}
\psi_{ij}^{(0)}=n_i\, n_j\ ,\qquad
\psi_{ij}^{(\pm)}=\frac{1}{2}\big(\delta_{ij} - n_i\, n_j\pm
i\epsilon_{ijk} n_k\big)\ . \label{psij}
\end{equation}
Similarly, the coefficients of $H_{\rm eff}$ can be calculated by
using  ${\cal
K}_{kl}^{(\alpha\beta)}(\xi\omega)$~\cite{Lindblad,pr5}.

\section{ Entanglement generation of accelerated atoms in presence of
a boundary}

With the basic formalism for the open system outlined in the
preceding section, we now start to examine whether entanglement can
be generated between two independent atoms moving with a constant
acceleration in the presence of a reflecting
 boundary, focusing our attention, in particular, on the difference between the
 accelerated atoms and static ones in the thermal bath at the corresponding Unruh temperature.

\subsection{Entanglement creation for the accelerated atoms aligned parallel
to the boundary }

\begin{figure}[htbp]
\centering
\includegraphics[scale=1]{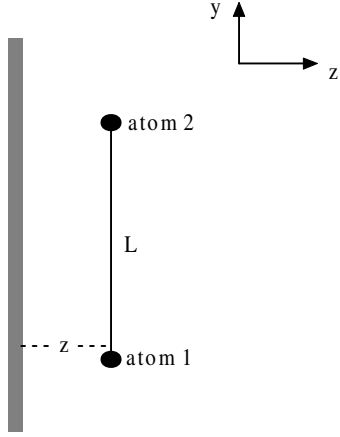}
\caption{ A reflecting plane boundary is located at z = 0 in space.
Two independent atoms separated from each other by a distance $L$
are aligned parallel to the boundary and accelerated in the
$x$-direction.}\label{pingxin}
\end{figure}

 For simplicity, we assume that two atoms are separated from each
other by a distance $L$ in $y-z$ plane (see Fig.~(\ref{pingxin})).
Furthermore,  both atoms are supposed to start moving at time $t=0$
with a constant proper acceleration $a$ in the $x$-direction and so
 their paths  can be described by
\begin{equation}\label{path}
x^0(t)=\frac{1}{a}\sinh(at),~x^1(t)=\frac{1}{a}\cosh(at)\;.
\end{equation}
with $x^2(t),x^3(t)$ components  remaining unchanged with their
initial values.

 Due to the assumption that the fields reflect from
  the boundary completely, we can use the  method of images to find the
  correlation functions (Eq.~(\ref{twogreen})),
\begin{eqnarray}\label{green-1}
G_{ij}^{(22)}(t-t')=G_{ij}^{(11)}(t-t')&=&-{1\over4\pi^2}
\bigg[{\delta_{ij}\over(x^0-{x^0}'-i\epsilon)^2-(x^1-{x^1}')^2-(x^2-{x^2}')^2-(x^3-{x^3}')^2}
\nonumber\\
 &&-{\delta_{ij}\over(x^0-{x^0}'-i\epsilon)^2-(x^1-{x^1}')^2-(x^2-{x^2}')^2-(x^3+{x^3}')^2}\bigg]\;\nonumber
 \\&=&-{\delta_{ij}\over16\pi^2}\bigg[{a^2\over\sinh^2[{a(t-t')\over2}-i\epsilon]}-{a^2\over\sinh^2[{a(t-t')
 \over2}-i \epsilon]-a^2z^2}\bigg]\;,
\end{eqnarray}
\begin{eqnarray}\label{green-2}
G_{ij}^{(21)}(t-t')=G_{ij}^{(12)}(t-t')&=&-{\delta_{ij}\over16\pi^2}
\bigg[{a^2\over\sinh^2[{a(t-t')\over2}-i\epsilon]-a^2L^2/4}\nonumber\\
&&-{a^2\over\sinh^2[{a(t-t')\over2}-i
\epsilon]-a^2z^2-a^2L^2/4}\bigg]\;.
\end{eqnarray}
 After some straightforward calculations, we can easily obtain their Fourier transforms
\begin{eqnarray}\label{g11g22}
&&{\cal G}^{(11)}_{ij}(\lambda)={\cal
G}^{(22)}_{ij}(\lambda)=\frac{\delta_{ij}}{2\pi}
\frac{\lambda}{1-e^{-2\pi \lambda/a}} -\frac{\delta_{ij}}{2\pi}
\frac{\lambda}{1-e^{-2\pi \lambda/a}}f_1(\lambda,z)\;,
\end{eqnarray}
\begin{eqnarray}\label{g12g21}
 {\cal G}^{(12)}_{ij}(\lambda)={\cal
G}^{(21)}_{ij}(\lambda)&=&\frac{\delta_{ij}}{2\pi}
\frac{\lambda}{1-e^{-2\pi
\lambda/a}}f_1(\lambda,L/2)-\frac{\delta_{ij}}{2\pi}
\frac{\lambda}{1-e^{-2\pi \lambda/a}}\nonumber\\&&\;\; \times
f_1\big(\lambda,\sqrt{z^2+L^2/4}\;\big)\;,
\end{eqnarray}
where $f_1(\lambda,z)$ is defined as
\begin{equation}
f_1(\lambda,z)=\frac{\sin[\frac{2\lambda}{a}\sinh^{-1}(az)]}{2z\sqrt{1+a^2z^2}\lambda}\;.
\end{equation}
 According to  Eq.~(\ref{cij-2}), we can write
\begin{eqnarray}\label{cabc}
&&C_{ij}^{(11)}=A_1 \delta_{ij}- iB_1
\epsilon_{ijk}n_k+C_1n_in_j\;,\nonumber\\
&&C_{ij}^{(22)}=A_2 \delta_{ij}- iB_2
\epsilon_{ijk}n_k+C_2n_in_j\;,\nonumber\\
&&C_{ij}^{(12)}=C_{ij}^{(21)}=A_3 \delta_{ij}- iB_3
\epsilon_{ijk}n_k+C_3n_in_j\;,
\end{eqnarray}
and the corresponding coefficients are
 \begin{equation}\label{A1A2}
A_1=A_2={\omega\coth(\pi\omega/a)\over4\pi}
 [1-f_1(\omega,z)]\;, B_1=B_2={\omega\over4\pi}[1-f_1(\omega,z)]\;,
 \end{equation}
\begin{equation}\label{c1c2}
C_1=C_2={a\over4\pi^2}[1-f_2(z)]-{\omega\coth(\omega\pi/a)\over4\pi}[1-f_1(\omega,z)]\;,
\end{equation}
\begin{equation}\label{A3}
A_3={\omega\cosh(\pi\omega/a)\over4\pi}\bigg[f_1(\omega,L/2)-f_1\big(\omega,\sqrt{z^2+L^2/4}\;\big)\bigg]\;,
\end{equation}
\begin{equation}\label{B3}
B_3=\frac{\omega}{4\pi}\bigg[f_1(\omega,L/2)-f_1\big(\omega,\sqrt{z^2+L^2/4}\;\big)\bigg]\;,
\end{equation}
\begin{eqnarray}\label{C3}
C_3&=&-{\omega\coth(\omega\pi/a)\over4\pi}\bigg[f_1(\omega,L/2)-f_1\big(\omega,\sqrt{z^2+L^2/4}\;\big)\bigg]\nonumber\\
&&+{a\over4\pi^2}\bigg[f_2(L/2)-f_2\big(\sqrt{z^2+L^2/4}\;\big)\bigg]\;.
\end{eqnarray}
 The new function $f_2(z)$ in the above expressions is given by
 \begin{equation}
 f_2(z)=\frac{\sinh^{-1}(az)}{z a\sqrt{1+a^2z^2}}\;.
 \end{equation}

 Similarly,  ${\cal K}_{ij}^{(\alpha\beta)}$ for the Hamiltonian $H_{\rm eff}$ can
 be obtained easily, but here we do not give the formulae in detail.
 As has already been discussed in detail elsewhere~\cite{Benatti1,Benatti2},
the effective Hamiltonian $H_{\rm
eff}=\tilde{H}_s^{(1)}+\tilde{H}_s^{(2)}+H_{\rm eff}^{(12)}$
includes three pieces. The first two correspond to the corrections
of the Lamb shift at a finite acceleration which should be
regularized according to the standard procedures in quantum field
theory and nevertheless they can be accounted for by replacing
$\omega$ the atom's Hamiltonian $H_s^{(1)}$ in Eq.~(\ref{hs}) with a
renormalized energy level spacing
\begin{equation}
\tilde{ \omega} =\omega +i [{\cal K}^{(11)}(-\omega)-{\cal
K}^{(11)}(\omega)]\;.
\end{equation}
Similarly, $\omega$  for  the  Hamiltonian $H_s^{(2)}$ is
\begin{equation}
\tilde{ \omega} =\omega +i [{\cal K}^{(22)}(-\omega)-{\cal
K}^{(22)}(\omega)]\;.
\end{equation}
 Meanwhile the third term is an environment generated direct
coupling between the atoms and is acceleration
independent\cite{Benatti1,Benatti2}. So the $H_{\rm eff}$ can be
ignored, since we are interested in the acceleration-induced
effects. Henceforth, we will only study the effects produced
 by the dissipative part ${\cal L}[\rho(t)]$.

 Using the explicit form of the master equation~(\ref{master}), we can
 investigate the time evolution of the reduced density  matrix and then we can figure out whether the
 state of the two-atom system is entangled or not  with the help of partial
 transposition  criterion~\cite{ppt}: a two-atom state $\rho(t)$ is entangled at $t$ if and only if
 the operation of partial transposition of $\rho(t)$ does not preserve its positivity.
 Let us now consider the
system in a finite time, and adopt a simple strategy for
ascertaining the entanglement creation at a neighborhood of the
initial time $t=0$, which has been introduced in Ref.~\cite{pr5}.
 For simplicity, we also let the initial pure, separable two-atom
 state be
 $\rho(0)=|+\rangle\langle+|\otimes|-\rangle\langle-|$ and consider
 the quantity
 \begin{equation}
{\cal Q}(t)=\langle\chi\vert\, \tilde{\rho}(t)\, \vert\chi\rangle\ ,
\end{equation}
where the tilde signifies partial transposition and $|\chi\rangle$
is a properly chosen $4$-dimensional vector. According to
Refs.~{\cite{Benatti2,pr5},  the entanglement of the system is
created  at the neighborhood of the time $t=0$ (i.e.,
$\partial_t{\cal Q}(0)< 0$), if and only if
\begin{equation}\label{condition}
\langle{u}|C^{(11)}|u\rangle\langle{v}|(C^{(22)})^T)|v\rangle<|\langle{u}|Re(C^{(12)})|v\rangle|^2\;,
\end{equation}
 where the subscript $T$ means matrix
transposition and the three-dimensional vectors $|u\rangle$ and
$|v\rangle$ can be chosen in a simple form as $u_i=v_i=\{1,-i,0\}$.
Using  Eq.~(\ref{cabc}) and taking the fact that $A_1=A_2$ into
account, we can compute Eq.~(\ref{condition}) for the vector
$\mathbf{n}$ along the third axis to get
\begin{equation}\label{bds-1}
{{A_3}^2\over {A_1}^2}+{{B_1}^2\over {A_1}^2}>1\;,
 \end{equation}
where
\begin{equation} \label{b1b2-1}
{{B_1}^2\over
{A_1}^2}=\bigg({1-e^{-2\pi{\omega}/a}\over1+e^{-2\pi{\omega}/a}}\bigg)^2=\bigg({1-e^{-{\omega}/T}\over1+e^{-{\omega}/T}}\bigg)^2
\;,
\end{equation}
with $T=a/2\pi$ being the Unruh temperature, and ${A_3}^2/{A_1}^2$
can be simplified as
\begin{equation}\label{FG-1}
{{A_3}^2\over {A_1}^2}=\frac{F}{G}\;,
\end{equation}
with
\begin{equation}\label{F1}
F=4\bigg\{ \frac{\sin\big[\frac{2\omega}{a}\sinh^{-1}\big(\frac{a
L}{2}\big)\big]}{L\omega\sqrt{4 +a^2L^2}}-
\frac{\sin\big[\frac{2\omega}{a}\sinh^{-1}\big({a\sqrt{L^2/4+z^2}}\;\big)\big]}{\omega\sqrt{L^2+4z^2}\sqrt{4
+a^2(L^2+4 z^2)}} \bigg\}^2\;
 \end{equation}
and
\begin{equation}\label{G1}
G=\bigg\{1- \frac{\sin[\frac{2\omega}{a}\sinh^{-1}(a z)]}{2z\omega
\sqrt{ 1+a^2z^2}}\bigg\}^2\;.
\end{equation}

 A comparison of Eq.~(\ref{bds-1}) with the condition for entanglement generation for
two independent static atoms immersed in a thermal bath with a
boundary (refer to Eq.~(23) in Ref.~\cite{zhjl}) reveals that the
uniformly accelerated atoms would, in general, behave differently
from the static atoms in a thermal bath at the Unruh temperature in
terms of entanglement creation at the neighborhood of time $t=0$,
since ${A_3}^2/{A_1}^2$ here is acceleration-dependent (or Unruh
temperature-dependent) whereas the corresponding term in Eq.~(23) in
Ref.~\cite{zhjl} does not rely on the bath temperature.

For a given atom, ${B_1}^2/ {A_1}^2$ is only dependent on the
acceleration $a$.
 When the acceleration is vanishingly
small, the value of ${B_1}^2/ {A_1}^2$ will approach to 1.
Therefore, inequality~(\ref{bds-1}) is always satisfied for zero
acceleration as long as  $L$ is not infinite.  On the other hand,
when the separation is vanishing ($L=0$), ${A_3}^2/{A_1}^2$ becomes
unity and inequality~(\ref{bds-1}) holds as long as the acceleration
is not infinite.

In order to figure out the difference in terms of the entanglement
generation between the accelerated atoms and the static ones in the
thermal bath at the Unruh temperature, let us recall that
${A_3}^2/{A_1}^2$ for the case of two static atoms in a thermal bath
with a boundary is given by \cite{zhjl}
\begin{equation}\label{asmall}
{{A_3}^2\over {A_1}^2}=\bigg[\frac{\sin(L
\omega)}{L\omega}-\frac{\sin(\omega\sqrt{L^2+4z^2})}{\omega\sqrt{L^2+4z^2}}\bigg]^2\bigg/
\bigg[1-\frac{\sin(2z\omega)}{2z\omega}\bigg]^2\;.
\end{equation}

First, we examine how the acceleration affects the entanglement
generation when the separation of the atoms  is comparable to the
characteristic wavelength of the atoms, i.e., when $L\sim 1/\omega$.
 In Fig.~(\ref{asmallz}),
 we plot  ${A_3}^2/{A_1}^2$ both for the case of static atoms in a thermal
bath (refer to Eq.~(\ref{asmall})) and that for the accelerated
atoms (refer to
 Eq.(\ref{FG-1})) as a functions of $aL$, with $L\omega$, the parameter
 characterizing the system of atoms fixed at
 $L\omega=1$ and  $z/L=(0.1,2,1000)$. The Figure shows that in general, ${A_3}^2/{A_1}^2$
 decreases as $aL$ or $a$ grows.   Furthermore, the smaller the value of  $z/L$
 is or the closer the system of the atoms is to the boundary, the
 more rapidly ${A_3}^2/{A_1}^2$ decays as a function of $aL$. Physically,
 this means that as the acceleration increases the atoms are less likely to get entangled,
 and the closer the two-atom system is to
the boundary, the more significantly the acceleration
 $a$ suppress  the possibility  of  entanglement production.
 Another conclusion one can draw from Fig.~(\ref{asmallz})
 is that the accelerated atoms are less likely to get entangled as
 compared to those static ones in the thermal bath at the Unruh
 temperature, since ${A_3}^2/{A_1}^2$ is always smaller for the accelerated
 atoms.

\begin{figure}[htbp]
\centering
\includegraphics[scale=0.8]{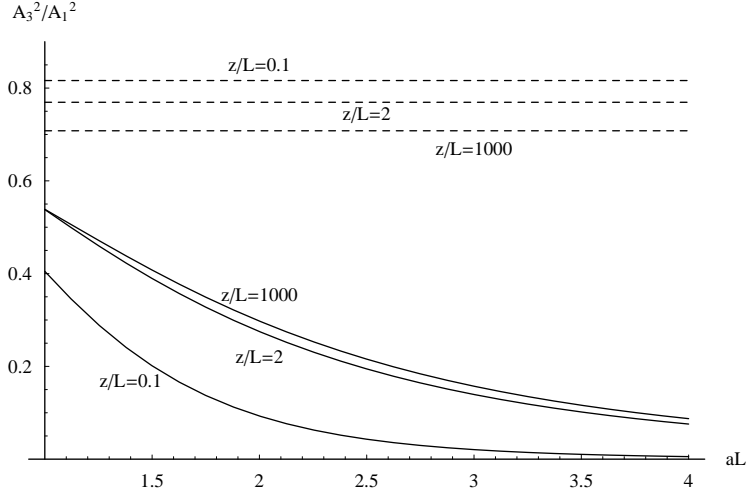}
\caption{ The solid lines denote  ${A_3}^2/{A_1}^2$ for the
accelerated atoms (Eq.~(\ref{FG-1})) as a function of $aL$  with
$L\omega=1$, and $z/L=(0.1,2,1000)$, and the dashed lines represent
that for the static atoms in a thermal bath ( Eq.~(\ref{asmall})).
}\label{asmallz}
\end{figure}

Now, let us discuss  how the acceleration affects the entanglement
generation when the two-atom system is at a distance from the
boundary comparable to the separation of the atoms, i.e., when
$z\sim L$. In Fig.~(\ref{asmallw}), we plot  ${A_3}^2/{A_1}^2$ as a
function of  $aL$ both for the case of static atoms in a thermal
bath (refer to Eq.~(\ref{asmall})) and that for the accelerated
atoms (refer to
 Eq.(\ref{FG-1})) with $z/L=1$ and $L\omega=(0.1,1,1.5,2.4)$. From
Fig.~(\ref{asmallw}), one can  see that  the value of
${A_3}^2/{A_1}^2$ for the accelerated atoms (Eq.~(\ref{FG-1})) is
obviously less than that for the static atoms in the Unruh thermal
bath (Eq.~(\ref{asmall}))  for the same value of $L\omega$, while
they both decrease rapidly with the increase of $L\omega$.
Thus, again, Fig.~(\ref{asmallw}) also shows  that  the entanglement
production is less likely to occur for accelerated atoms than the
static ones immersed in a thermal bath at the corresponding ``Unruh
temperature".  So, in terms of entanglement generation at at a
neighborhood of the initial time $t=0$, independent accelerated
atoms do not behave as if they were static in a thermal bath at the
Unruh temperature.

\begin{figure}[htbp]
\centering
\includegraphics[scale=0.8]{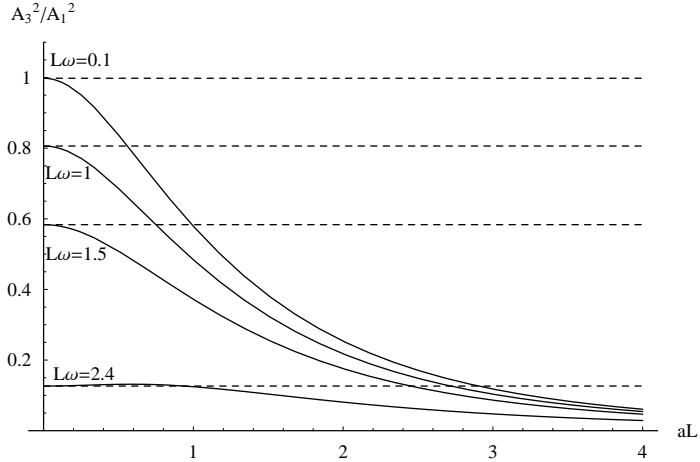}
\caption{ ${A_3}^2/{A_1}^2$ for accelerated atoms as function of
$aL$ is described by solid lines with $z/L=1$ and $L\omega=0.1,\;
1,\; 1.5,\; 2.4$, respectively,  and that for the static atoms in a
thermal bath by the dashed lines. }\label{asmallw}
\end{figure}

At this point, one may wonder what the effects are of the presence
of the boundary on entanglement generation of the two-atom system.
To address this issue, let us analyze and compare two special cases.
One is when the atom system is very close to the boundary, i.e.,
when $z\ll{L}$, $za\ll1$ and $z\omega\ll1$. In this case,
${A_3}^2/{A_1}^2$ (Eq.~(\ref{FG-1})) can be approximated as
\begin{eqnarray}\label{zsmall-1}
{A_3}^2/
{A_1}^2&\approx&\frac{144}{L^6(4+a^2L^2)^2\omega^2(a^2+\omega^2)^2}
\bigg\{\frac{2+a^2L^2}{\sqrt{4+a^2L^2}}\sin\bigg[\frac{2\omega}{a}\sinh^{-1}(a
L/2)\bigg]
\nonumber\\&&-L\omega\cos\bigg[\frac{2\omega}{a}\sinh^{-1}(a
L/2)\bigg]\bigg\}^2+O\bigg(\frac{z^2}{L^2}\bigg)\;.
\end{eqnarray}
The other case is when  the system is located  very far away from
the boundary (i.e., $z\gg{L}$, $az\gg1$ and $\omega{z}\gg1$). One
then has
\begin{equation}\label{zlarge}
\frac{{A_3}^2}{{A_1}^2}=\frac{4
\sin^2[\frac{2\omega}{a}\sinh^{-1}(\frac{a
L}{2})]}{L^2(4+a^2L^2)\omega^2}+ O\bigg(\frac{L^2}{{z}^2}\bigg)\;.
\end{equation}
In the Fig.~(\ref{wL}), we have plotted ${A_3}^2/{A_1}^2$ as a
function of $L\omega$ for these special cases with the value of $aL$
chosen as 0.2, 1, and 3. Plot (a) is basically the same as that of
Fig.~(2) in Ref.~\cite{zhjl}.  So,  when the acceleration is very
small ($a\ll 1/L$), the effect of the presence of a boundary on the
entanglement generation of the accelerated atoms is essentially the
same as that of the static atoms in the thermal bath at the Unruh
temperature, i.e., when $L\omega$ is  small, approximately smaller
than four, the presence of the boundary may make the accelerated
atoms be entangled which would otherwise still be separable, since
${A_3}^2/{A_1}^2$ is always greater with presence of the boundary
than that without.  However, plot (c) shows that when the
acceleration is very large ($a\gg 1/L$), ${A_3}^2/{A_1}^2$ is always
smaller with presence of the boundary than that without, thus the
entanglement production is more likely for the system located
farther away from the boundary no matter what the value of $L\omega$
is. Meanwhile, when the acceleration is comparable to the separation
of the atoms, then one finds from plot (b) that ${A_3}^2/{A_1}^2$ is
smaller with the presence of the boundary than without when
$L\omega$ is approximately less than two and larger than four,
whereas it is greater when $L\omega$ is approximately in  between
two and four, suggesting that for a given kind of atom, the
separation will have significant influence on whether the
entanglement will be created and the presence of the boundary may
make the accelerated atoms entangled which would otherwise still be
separable or vice versa, depending crucially on the distance between
the atoms.  Therefore, when the acceleration is not small as
compared to the separation of the atoms, the accelerated atoms
exhibit distinct features from the static ones in the thermal bath
at the Unruh temperature when entanglement generation is concerned.

\begin{figure}[htbp]
\centering
\includegraphics[height=2.1in,width=2.1in]{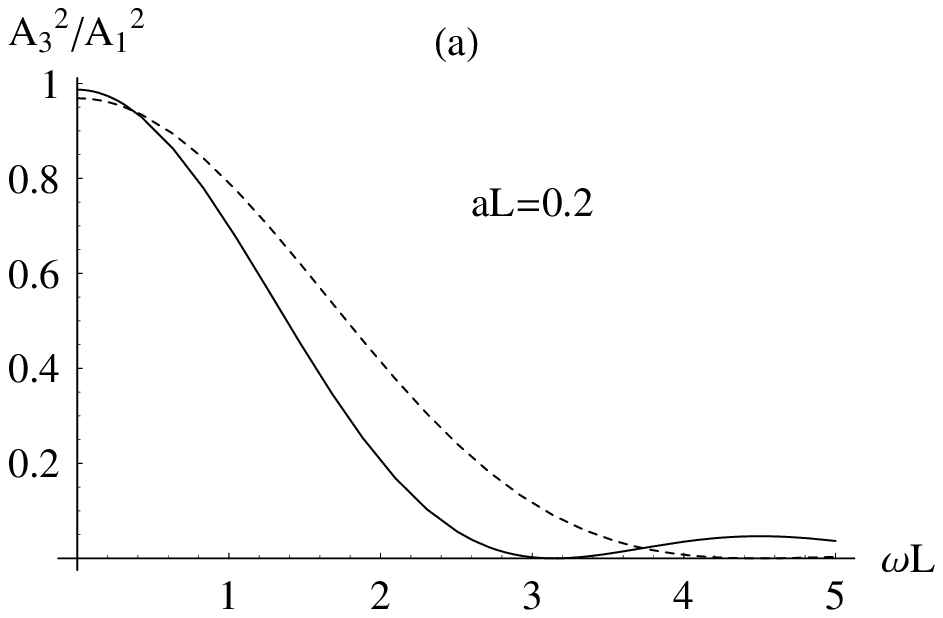}\includegraphics[height=2.1in,width=2.1in]{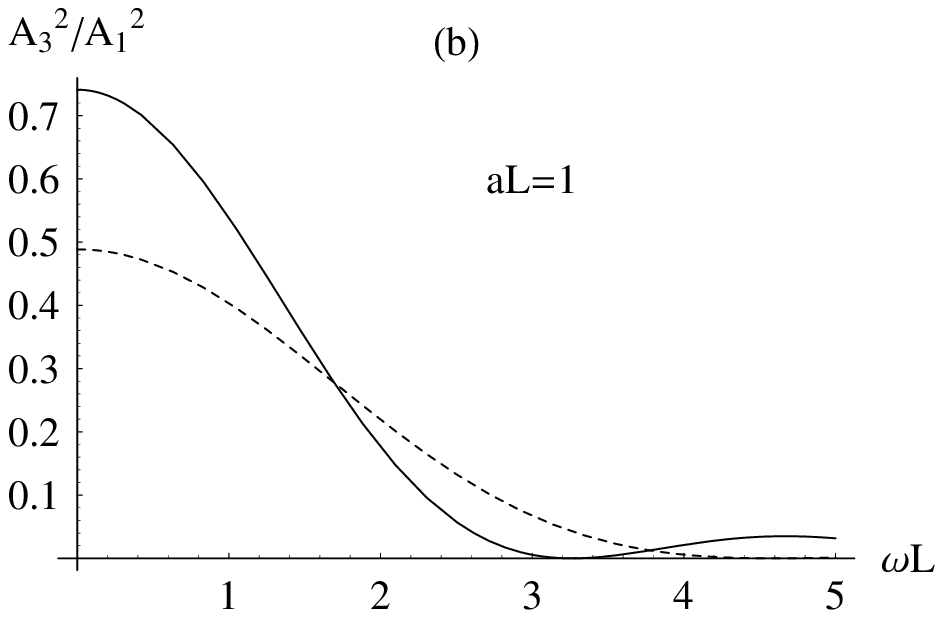}
\includegraphics[height=2.1in,width=2.1in]{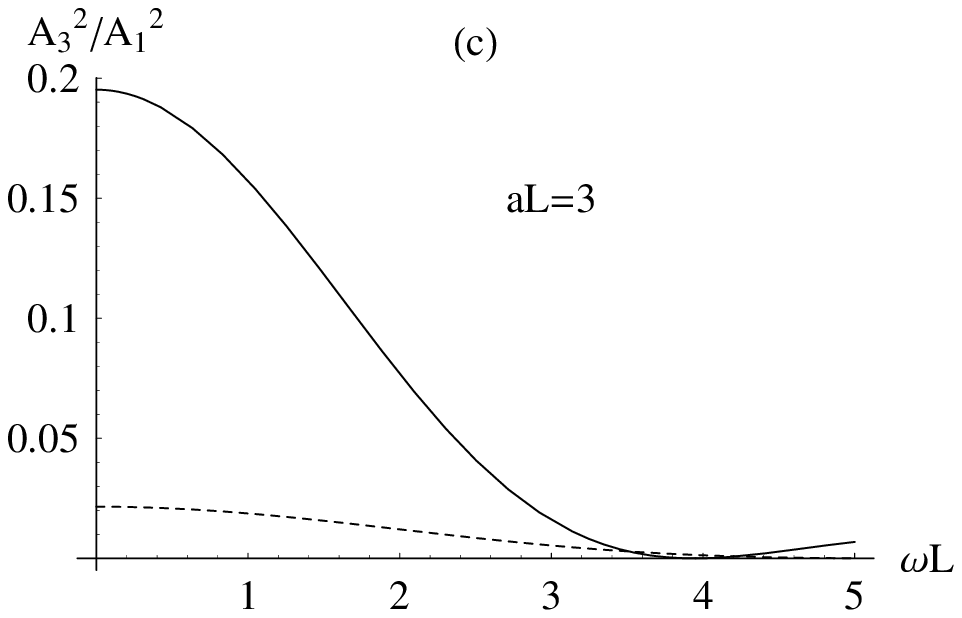}
\caption{ The dashed lines denote ${A_3}^2/{A_1}^2$ as function of
$\omega L$ with $aL=\{0.2,1,3\}$, for vanishingly small $z$ (see
Eq.(\ref{zsmall-1})), and the solid lines describe that for $z$
approaching infinity(see Eq.~(\ref{zlarge})). }\label{wL}
\end{figure}

\subsection{The entanglement creation for the accelerated atoms vertically aligned to the boundary plane }

Let us now briefly examine another special alignment of the two-atom
system, that is,  the case in which the two-atom system is
vertically aligned to the boundary(see Fig.~(\ref{chuizhi})).
\begin{figure}[htbp]
\centering
\includegraphics[scale=1]{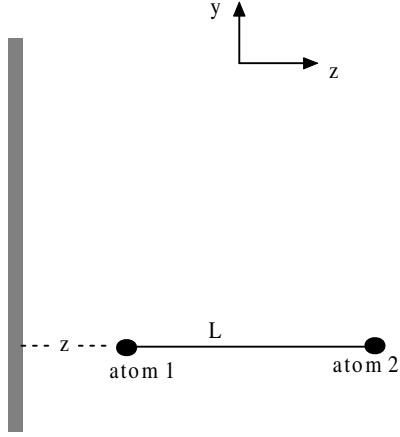}
\caption{ The atom system is aligned vertical to the boundary and
$z$ is the distance between the boundary and the atom which is
closer. The acceleration is again in the
$x$-direction.}\label{chuizhi}
\end{figure}
Now one finds in the correlation functions of the scalar fields that
$G_{ij}^{(22)}(t-t')\neq{G_{ij}^{(11)}(t-t')}$ as a result of the
unequal distance from the boundary for the two atoms. Therefore, we
have
\begin{eqnarray}\label{g11g22-2}
&&{\cal G}^{(11)}_{ij}(\lambda)=\frac{\delta_{ij}}{2\pi}
\frac{\lambda}{1-e^{-2\pi \lambda/a}} -\frac{\delta_{ij}}{2\pi}
\frac{\lambda}{1-e^{-2\pi \lambda/a}}f_1(\lambda,z)\;,
\nonumber\\
&&{\cal G}^{(22)}_{ij}(\lambda)=\frac{\delta_{ij}}{2\pi}
\frac{\lambda}{1-e^{-2\pi \lambda/a}} -\frac{\delta_{ij}}{2\pi}
\frac{\lambda}{1-e^{-2\pi \lambda/a}}f_1(\lambda,z+L)\;,
\end{eqnarray}
\begin{eqnarray}\label{g12g21-2}
 {\cal G}^{(12)}_{ij}(\lambda)={\cal
G}^{(21)}_{ij}(\lambda)&=&\frac{\delta_{ij}}{2\pi}
\frac{\lambda}{1-e^{-2\pi
\lambda/a}}f_1(\lambda,L/2)-\frac{\delta_{ij}}{2\pi}
\frac{\lambda}{1-e^{-2\pi \lambda/a}}\nonumber\\&&\;\; \times
f_1\big(\lambda,z+L/2\;\big)\;.
\end{eqnarray}

Similarly, using Eq.~(\ref{cij-2}) and Eq.~(\ref{cabc}), we can also
write
\begin{equation}\label{A1A2-2}
 A_1={\omega\coth(\pi\omega/a)\over4\pi}
 [1-f_1(\omega,z)]\;,\;\;
A_2={\omega\coth(\pi\omega/a)\over4\pi}
 [1-f_1(\omega,z+L)]\;,
 \end{equation}
\begin{equation}
 B_1={\omega\over4\pi}[1-f_1(\omega,z)]\;, B_2={\omega\over4\pi}[1-f_1(\omega,z+L)]\;,
 \end{equation}
\begin{eqnarray}\label{c1c2-2}
C_1={a\over4\pi^2}[1-f_2(z)]-{\omega\coth(\omega\pi/a)\over4\pi}[1-f_1(\omega,z)]\;,\nonumber\\
C_2={a\over4\pi^2}[1-f_2(z+L)]-{\omega\coth(\omega\pi/a)\over4\pi}[1-f_1(\omega,z+L)]\;,
\end{eqnarray}
\begin{equation}\label{A3-2}
A_3={\omega\cosh(\pi\omega/a)\over4\pi}\bigg[f_1(\omega,L/2)-f_1\big(\omega,z+L/2\;\big)\bigg]\;,
\end{equation}
\begin{equation}\label{B3-2}
B_3=\frac{\omega}{4\pi}\bigg[f_1(\omega,L/2)-f_1\big(\omega,z+L/2\;\big)\bigg]\;,
\end{equation}
\begin{eqnarray}\label{C3-2}
C_3&=&-{\omega\coth(\omega\pi/a)\over4\pi}\bigg[f_1(\omega,L/2)-f_1\big(\omega,z+L/2\;\big)\bigg]\nonumber\\
&&+{a\over4\pi^2}\bigg[f_2(L/2)-f_2\big(z+L/2\;\big)\bigg]\;.
\end{eqnarray}
with the above results, one can show that the condition for
entanglement creation, Eq.(\ref{condition}), becomes
 \begin{equation}\label{bds-2}
{{A_3}^2\over {A_1A_2}}+{B_1B_2\over {A_1A_2}}>1\;,
 \end{equation}
where
\begin{equation} \label{b1b2-2}
{B_1B_2\over
{A_1A_2}}=\bigg({1-e^{-2\pi{\omega}/a}\over1+e^{-2\pi{\omega}/a}}\bigg)^2=\bigg({1-e^{-{\omega}/T}\over1+e^{-{\omega}/T}}\bigg)^2
\;,
\end{equation}
with $T=a/2\pi$  also being the Unruh temperature.
Eq.~(\ref{b1b2-2}) is the same as Eq.~(\ref{b1b2-1}) and is only
dependent on the parameters $a$ and $\omega$. Therefore we only need
to discuss the first term in Eq.~(\ref{bds-2}), which can be shown
to be given by
 \begin{equation}\label{FG-2}
{{A_3}^2\over {A_1A_2}}=\frac{F'}{G'}\;,
\end{equation}
where
\begin{equation}\label{F2} F'=4\Bigg\{\frac{\sin\big[\frac{2
\omega}{a}\sinh^{-1}\big(\frac{a
L}{2}\big)\big]}{L\omega\sqrt{4+a^2L^2}}-\frac{\sin\big[\frac{2\omega}{a}\sinh^{-1}\big(\frac{a
{L}+2 a{z}}{2}\big)\big]} {(L+2 z)\omega\sqrt{4+a^2(L+2
z)^2}}\Bigg\}^2\;,
\end{equation}
\begin{equation}\label{G2}
G'=\bigg[1-
 \frac{\sin[\frac{2\omega}{a}\sinh^{-1}(a
z)]}{2z\omega\sqrt{1+a^2z^2}}\bigg]\bigg[1-\frac{\sin[\frac{2\omega}{a}\sinh^{-1}(aL+a
z)]}{2\omega(L+z)\sqrt{1+a^2(L+z)^2}}\bigg]\;.
\end{equation}
If we expand Eq.~(\ref{FG-2}) in the limit of $z\rightarrow\infty$,
  we find that the result is the same as that of
Eq.~(\ref{zlarge}). This is  consistent with what one expects, since
very far from the boundary, the space should be almost isotropic.
Thus, as an example to demonstrate the difference, in terms of the
entanglement generation, between case when the atom system is
aligned parallel to boundary and that when it is vertically aligned,
we will analyze the situation when the atom is located very close to
the boundary. In the limit of vanishingly small $z$,
${A_3}^2/({A_1A_2})$ reads
\begin{eqnarray}\label{zsmall-2}
{{A_3}^2\over
({A_1A_2})}\approx&&\frac{\bigg\{L\omega\sqrt{a^2L^2+4}\cos\big[\frac{2\omega}{a}\sinh^{-1}\big(\frac{a
L}{2}\big)\big]-(a^2L^2+2)
\sin\big[\frac{2\omega}{a}\sinh^{-1}\big(\frac{a
L}{2}\big)\big]\bigg\}^2}{2L\omega\sqrt{1+a^2L^2}
-\sin\big[\frac{2\omega}{a}\sinh^{-1}\big(a L\big)\big]}\times
\nonumber\\
&&\frac{192\;\sqrt{1+a^2L^2}}{L^3\omega(4+a^2L^2)^3(a^2+\omega^2)}+O
\bigg(\frac{z}{L}\bigg)\;.
\end{eqnarray}
In Fig.~(\ref{czh-px}), we have plotted  ${A_3}^2/({A_1A_2})$ as
function of $L\omega$ with $aL=\{0.2,1,5\}$, both for the parallel
two-atom system  and the vertical one, when  $z$ is vanishingly
small. Notice that for the parallel system $A_2=A_1$. These plots in
the Figure shows clearly that generically  the value of
${A_3}^2/({A_1A_2})$ for the parallel aligned atom system is smaller
than the vertically aligned one.

\begin{figure}[htbp]
\centering
\includegraphics[height=2.1in,width=2.1in]{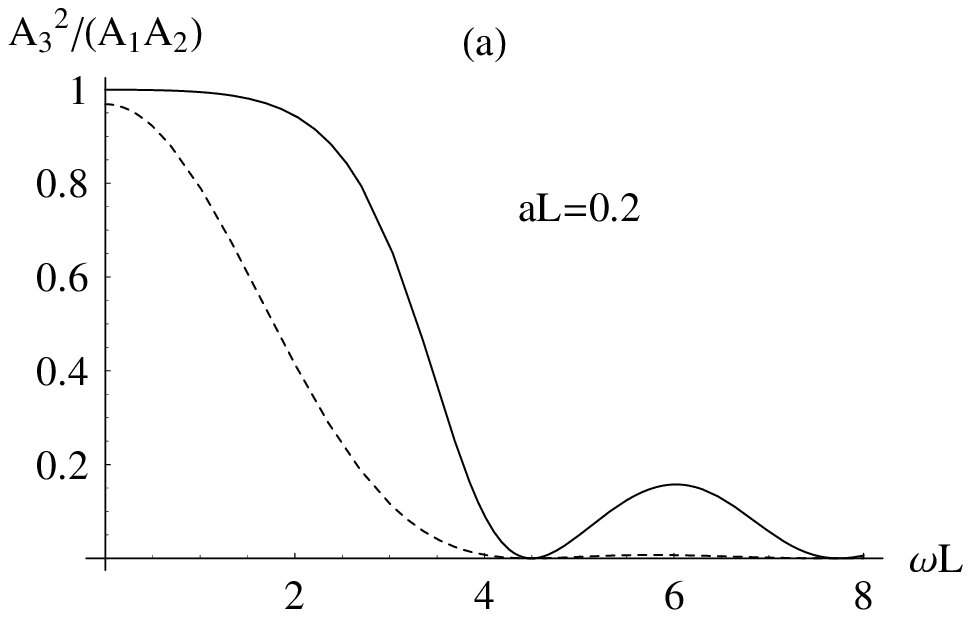}\includegraphics[height=2.1in,width=2.1in]{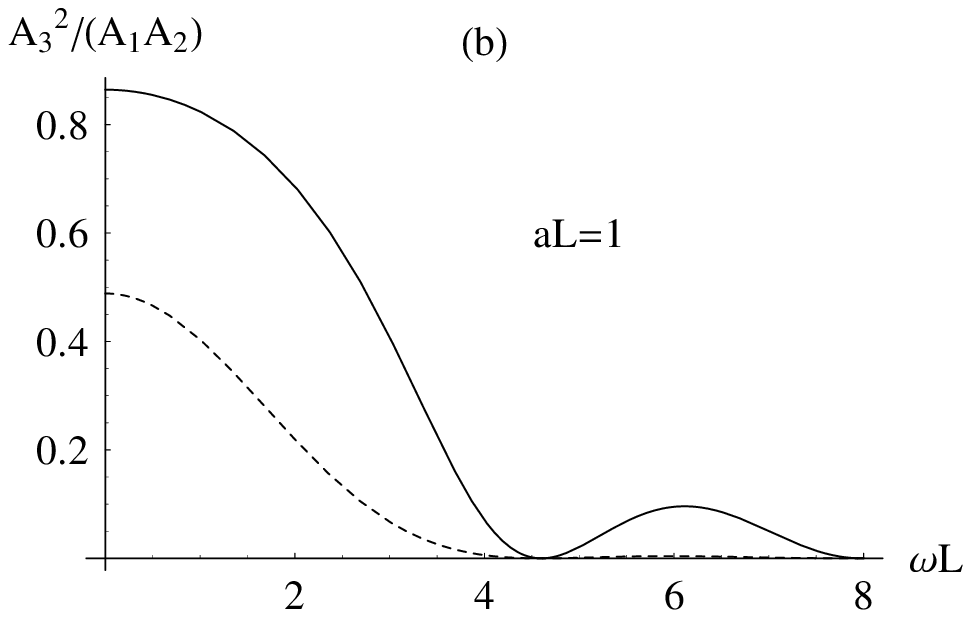}
\includegraphics[height=2.1in,width=2.1in]{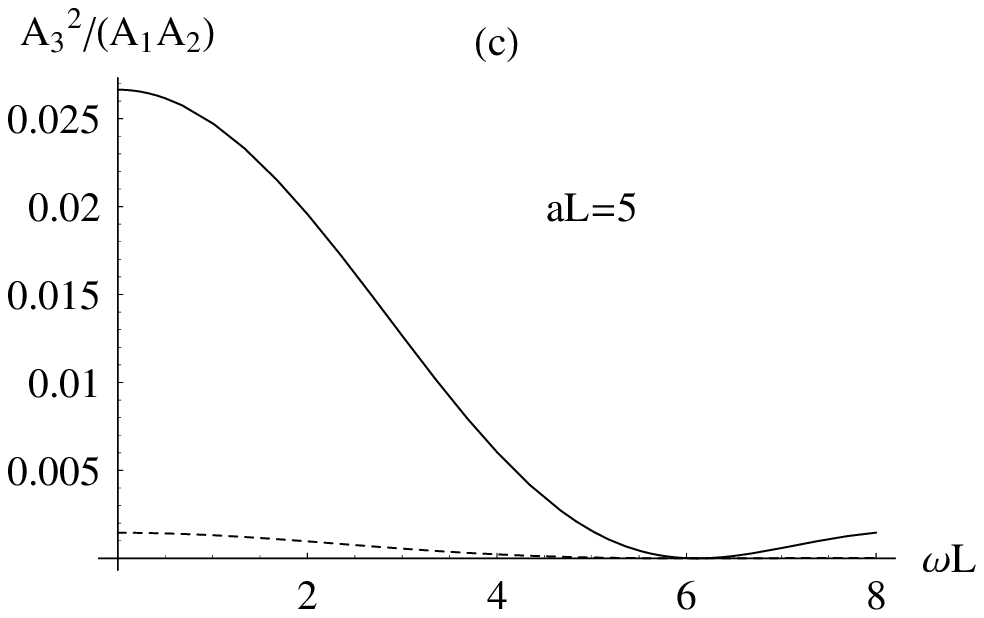}
\caption{ The dashed lines denote ${A_3}^2/{A_1}^2$ as function of
$\omega L$ with $aL=\{0.2,1,5\}$, for vanishingly small $z$ when the
atoms are aligned parallel (see Eq.(\ref{zsmall-1})), and the solid
lines describe that when the atoms are aligned vertically (see
Eq.~(\ref{zsmall-2})). }\label{czh-px}
\end{figure}
Therefore, we can conclude that very close to boundary, accelerated
atoms that are  aligned parallel to the boundary are less likely to
get entangled than those that are vertically aligned.

\section{Conclusion and discussion}

Using the open system  paradigm, we have investigated, at a
neighborhood of the initial time, the entanglement generation of
independent uniformly accelerated atoms interacting with scalar
fields in vacuum  with the presence of a reflecting plane boundary.

Our results reveal that, for the parallel two-atom system,  both
when the separation of the atoms is comparable to the characteristic
wavelength of the atoms (but the atom system is not very close to
the boundary), i.e., when $L\sim 1/\omega$, and  when the two-atom
system is at a distance from the boundary comparable to the
separation of the atoms, i.e., when $z\sim L$,  the entanglement
production is less likely to occur for accelerated atoms than the
static ones immersed in a thermal bath at the corresponding ``Unruh
temperature".

On the other hand, if the atom system is very close to the boundary,
i.e., if $z\ll{L}$, $za\ll1$ and $z\omega\ll1$, then  when the
acceleration is very large ($a\gg 1/L$), the presence of the
boundary will always make the entanglement production less likely to
happen no matter what the value of $L\omega$ is. Meanwhile, when the
acceleration is comparable to the separation of the atoms, then the
accelerated atoms are less likely to get entangled with the presence
of the boundary than without when $L\omega$ is approximately less
than two and larger than four, whereas they are likely to do so when
$L\omega$ is approximately in between two and four, suggesting that
for a given kind of atom, the separation between the atoms will have
significant influence on whether the entanglement will be created
and the presence of the boundary may make the accelerated atoms
entangled which would otherwise still be separable or vice versa.
This is in sharp contrast to the static atoms in thermal bath where
it has been shown that when $L\omega$ is small, approximately
smaller than four, the presence of the boundary may make the atoms
be entangled which would otherwise still be separable and only when
$L\omega$ is very large, will the presence of the boundary always
make the entanglement production less likely to occur~\cite{zhjl}.

Therefore, in terms of the entanglement generation at a neighborhood
of the initial time,  the accelerated atoms exhibit distinct
features from the static ones in a thermal bath at the Unruh
temperature. In other words, accelerated atoms in vacuum do not have
to behave as if they were static in a thermal bath.  A similar
example of this kind is that a uniformly accelerated proton does not
have to behave as if it were static in a thermal bath at the Unruh
temperature in terms of its lifetime against weak decay~\cite{MV}.

\begin{acknowledgments}
 This work was supported in part  by the National Natural Science
Foundation of China  under Grant No.10575035, the Program for New
Century Excellent Talents in University (NCET, No. 04-0784), the Key
Project of Chinese Ministry of Education (No. 205110), and the
National Basic Research Program of China under Grant No.
2003CB71630.
\end{acknowledgments}

\end{document}